\documentstyle[pra,aps,epsfig]{revtex}
\def\part#1#2{{\partial#1\over\partial#2}}
\def\ggg {{\cal G}}

\def\hhh {{\cal H}}

\def\jpa#1#2#3 {J. Phys. A: Math. Gen. {\bf{#1}}, #2 (19#3)}

\def\textit#1{{\it #1}}
\def\be{\begin{equation}}
\def\ee{\end{equation}}
\def\textbf#1{{\bf #1}}
\def\textsl#1{{\sl #1}}
\def\emph#1{{\textit#1}}
\def\mathbf#1{{\bf#1}}
\def\mathcal#1{{\cal#1}}

\begin{document}
\draft
\flushbottom
\twocolumn[
\hsize\textwidth\columnwidth\hsize\csname @twocolumnfalse\endcsname

\title{STRUCTURAL INVARIANCE AND THE ENERGY SPECTRUM}
\author{F. LEYVRAZ, R. A. MENDEZ AND T.H. SELIGMAN}
\address{Centro de Ciencias F\'{\i}sicas, UNAM. \\
Av. Universidad s/n. Col. Chamilpa \\
62210 Cuernavaca, Mor., MEXICO }
\maketitle

\tightenlines
\widetext
\advance\leftskip by 57pt
\advance\rightskip by 57pt

\begin{abstract}
\ We extend the application of the concept of structural
invariance to bounded time independent systems. This concept, previously
introduced by two of us to argue that the connection between random matrix
theory and quantum systems with a chaotic classical counterpart is in fact
largely exact in the semiclassical limit is extended to the energy spectra
of bounded time independent systems. We proceed by showing that the results
obtained previously for the quasi-energies and eigenphases of the S-matrix
can be extended to the eigenphases of the quantum Poincar\'{e} map which is
unitary in the semiclassical limit. We then show that its eigenphases in the
chaotic case move rather stiffly around the unit circle and thus their local
statistical fluctuations transfer to the energy spectrum via Bogomolny's
prescription. We verify our results by studying numerically the properties
of the eigenphases of the quantum Poincar\'{e} map for billiards by using
the boundary integral method.
\end{abstract}

]

\narrowtext
\tightenlines
\setcounter{equation}{0}

\newpage
\section{Introduction}

In previous papers two of us developed the concept of structural invariance
for periodically driven \cite{struct} as well as for scattering systems 
\cite{goslar,cas-chir} in order to establish the connection between random
matrix theory on the one hand and quantum systems with a chaotic classical
counterpart on the other. Basically we recovered for such systems the result
predicted for the long-range behaviour of the spectral two-point function in
a seminal paper by Berry \cite{berry} and demonstrated numerically 
and experimentally in a large, and growing, number of examples
\cite{oriol,thomas}: Namely, the fluctuation properties of
quantum systems whose classical counterpart is chaotic behave as those of
an appropiately chosen ensemble of random matrices.
There have been numerous attempts to account
for this connection. Among the most notable are periodic orbit theory,
which shows how certain random matrix properties arise
from a semiclassical formula connecting the density of states
with a given sum over periodic orbits. Another approach, which
has achieved some remarkable successes recently, is the
supersymmetric approach, which has been able to show how 
such phenomena arise in the context of disordered
systems \cite{Efetov}. 
For chaotic systems without disorder, however, the
results obtained in this manner are still far from
convincing \cite{Altshuler,comment}. 
The approach taken by structural invariance
is in some sense different, since it takes an
explicitly probabilistic point of view, asserting only that 
a chaotic system has RMT properties with probability
one and only speaking in terms of ensembles of systems.
\cite{struct,wiko,Guadalajara}. This approach has already
led to unexpected predictions, which since
have been confirmed numerically \cite{Charles}
and later by arguments taken from periodic orbit 
theory \cite{Keating&Robbins}.
            
A weak point of our approach was the fact that its application was
in principle restricted to the discussion of eigenphase statistics. 
In this paper, we shall
use the semiclassical quantization of the Poincar\'{e} map and
its relation to the energy spectrum in order to transfer the
statistics of eigenphases to that of eigenvalues. 
The following relation, due to Bogomolny \cite{bog} 
\begin{equation}
\det (1-T(E_n))=0
\end{equation}
shows that we obtain an energy eigenvalue $E_n$ each time one of the
eigenphases of the matrix $T$ (operator associated to the quantum
Poincar\'{e} map) takes the value 
unity; thus a stiff rotation of the unit
circle on which the eigenphases are located would transfer their spectral
fluctuations to those of the energy eigenvalues.
In this paper we shall show that the assumption of fairly uniform
speed of the eigenphases as a function of energy is an excellent
approximation for chaotic Poincar\'{e} maps.

The structure of this paper is as follows: First, in Section 2,
we give a short account of structural invariance in the most
general terms possible. In Section 3, we review how the connection
between classical behaviour and quantum statistics is effected in the 
case of canonical maps. In Section 4, we proceed to generalize the analysis to
Poincar\'e maps. There we show, using Bogomolny's semiclassical 
quantization, that the spectral statistics of the eigenphases 
of the Poincar\'e map essentially carry over to those of the eigenvalues.
In Section 5, we illustrate the results by means of a numerical 
computation of the stadium billiard.
\section{Structural invariance}

The point we want to discuss in this Section
concerns the construction of a reasonable ensemble given an
individual system. Such problems are not really new to physics: indeed,
whenever one attempts to describe an individual system by an ensemble, as
occurs for example in statistical mechanics, the problem of deriving the
ensemble from the individual system arises. In such cases it is well known
that one first needs to identify a set of relevant properties. Once this has
been done, we define the structural invariance group of the object as the
group of all transformations which transform the given object into one with
the same relevant properties. 

First, let us present a general method to derive such an ensemble.
Let us consider
a set of objects with given ``relevant''properties---the
choice of these will turn out to be crucial, though it is usually dictated
from the problem at hand---and a
specific object $A$ taken from this set. 
Furthermore, let the various objects be transformed
among each other through the action of a group $\ggg$.
The object $A$ has a certain number of the ``relevant'' properties.
We define $\hhh$, {\it the structural invariance group of $A$\/},
in the following manner: Let $\hhh$ be that subgroup of $\ggg$
which leaves all properties of $A$ invariant. If we now act
on $A$ with all the transformations of $\hhh$, we generate a collection
$\Omega$ of objects, which has as additional structure that 
of a homogeneous space under the group $\hhh$. If $\hhh$ additionally
has an invariant measure, then a measure on $\Omega$ can be generated in 
a natural way as follows: To each subset $\Sigma$
of $\Omega$ one can associate the set of transformations $\cal T$
in $\hhh$ which map $A$ in $\Sigma$. The Haar measure of $\cal T$
can then be taken as the measure of $\Sigma$. This is 
independent of the choice of $A$, because of the
invariance of the Haar measure. Furthermore, it is a
measure on $\Omega$ that is invariant under the group
action and it is therefore clearly singled out. 

Let us first consider a trivial example: Consider the set of all 
infinite binary sequences of ones
and minus ones. As relevant properties, we take the 
value of all finite range correlation functions. By this we mean the 
following: Consider the sequence $(s_i)$. The two-point
correlation is then given by
\be
C_i=\lim_{N\to\infty}{1\over2N}\sum_{k=-N}^Ns_ks_{k+i}
\ee
and more general correlations defined through the obvious generalization.
As transformations acting on binary sequences, we consider
all possible transformations. However, the resulting group is
highly pathological and has presumably no Haar measure. One way to
proceed is to limit ourselves to large but finite sequences. 
The group $\ggg$ then consists of all transformations 
mapping arbitrary finite sequences on other arbitrary finite
sequences.
If the correlations of the original sequence are
sufficiently small, then this property
is respected by almost all transformations
of $\ggg$, that is all but a set of very small
measure. The structural invariance
group $\hhh$ is then the full group. From this follows that we can consider
the original sequence as a representative element of the full ensemble
of all binary sequences, in which all sequences are taken to be
equally probable. On the other hand, if there is a clear predominance 
of ones, say, but no further correlations, then we can take $\hhh$
to be the set of all maps which permute the elements of the original
sequence. The set $\Omega$ is then the set of all sequences which arise
from the original one through permutation, all sequences being equally
probable. Again, it is reasonable to assume that the original 
sequence is indeed a representative member of the ensemble. In this
sense, whatever property holds with probability one in the
ensemble that has been constructed in this fashion will also 
presumably hold for the original element. Note that this approach
does not always work: For example, if I find a non-zero
two-point correlation, it is at least not evident how to define
a group $\hhh$ that respects this property. A better approach
might then be to define a {\it non-invariant\/} measure on 
$\ggg$ that takes this correlation into account and that respects
it with probability one. However, it will turn out that we 
shall be able to define a structural invariance group 
in the majority of the cases we shall be dealing with.

\section{Canonical Maps: Asociated Quantum and Classical Ensembles}

We now wish to apply the above general approach to the particular
case of bijective canonical maps on a compact phase space. 
To this end, we must define
the various ingredients involved in the construction. The set of
all objects is the set of all bijective canonical transformations
on a compact phase space $\Gamma$. The specific object we start from 
is a given such transformation $C_0$. 
As for the ``relevant'' properties, their choice is dictated
by the nature of the problem. In the followoing, 
we shall principally be interested in the short-range 
statistical properties of the eigenphases and eigenvalues
of the associated unitary and hermitian operators. 
This corresponds to the long-time properties 
of the classical time evolution. For this reason,
we shall consider only such properties 
to be ``relevant'' as remain invariant under arbitrary iteration
of the map. As for the group $\ggg$, we shall take it to be the
direct product of the group of all bijective canonical canonical
transformations with itself, with the following action on the set 
of all bijective canonical transformations:
\be
p_{C_\alpha,C_\beta}:C\to C_\alpha CC_\beta.
\ee
To complete the construction, it would be necessary to have a Haar measure
on the group $\ggg$. As far as we know, the existence of such a measure
has neither been proved nor shown to be impossible. While we
believe that such a measure does indeed exist in the case of compact
phase spaces, we shall not make siuch an
assumption, but rather make the following
consideration: After all, what we are really 
interested in are not the 
(classical) properties of the canonical maps, but rather those
of the corresponding quantum unitary maps. We therefore need to quantize
both $C_0$ and the set of all canonical transformations. To this
end, we first need to know the quantum equivalent of the phase 
space $\Gamma$. Since $\Gamma$ is compact, its quantization will be
a finite dimensional Hilbert space, with the dimension
$N$  going to infinity as
$\hbar\to0$. Note that in general not all values of $\hbar$
yield a quantization of $\Gamma$, but only such for which 
cells of size $(2\pi\hbar)^d$ fit in an integer nuber of times
in the phase space $\Gamma$. We now proceed to quantize all canonical maps
using some quantization technique. Two things shoud be noted
here: First, there is no unique way to quantize a canonical map. Second, 
whichever way is used to associate a unitary map to a canonical map,
it can never be such that the quantization of the produvct
of two canonical maps is the product of the two quantized
maps. Such a result can in deed be obtained, but only approximately,
in the limit $N\to\infty$. After having constructed
$\Omega$ in the set of all classical maps, we can proceed to define
$\Omega_Q$ to consist of the set of all unitary maps corresponding to a
given quantization of the maps in $\Omega$.
In a similar way, we 
can translate the group $\ggg$ in the quantum domain, where it becomes
simply $U(N)\times U(N)$, where $N$ is the dimension of the Hilbert
space. The finiteness of $N$ is a crucial point here, and this is
precisely the point at which we use the fact that the phase space $\Gamma$
is compact. From this follows that there exists a (unique)
Haar measure on the quantized version of $\ggg$ and hence, for every
closed subgroup $\hhh$ as well. 

There is an apparent problem with the above program, however: 
Since the ``translation'' from classical to quantum language is by no
means unique, one cannot assign to each classical 
canonical map a well-defined unitary map. Nevertheless, since we
are only interested in the semiclassical limit, this problem 
is not really severe, since all different possible choices
will be very close to each other. One might worry, further, that 
an approximate translation of the group $\hhh$ from the classical 
to the quantum domain might lead one to lose the group property. 
But this is not the case, since the group $\hhh$ is defined 
through the invariance of the properties of $C_0$. This definition
can itself be carried over to the quantum domain and the group
property is then trivially maintained.

Another point that can be raised is the following: Most canonical
maps of interest are ``simple'' or smooth, whereas there is no
reason to believe that such smoothness is characteristic of 
``arbitrary'' canonical maps, if this can be given a meaning. 
Certainly, for unitary maps, it is easy to see that arbitrary
unitary maps in a (high-dimensional) Hilbert space representing
the semiclassical counterpart of a given classical system with
probability one fail to correspond to any reasonable classical
canonical transformation. In the light of these facts, one might worry that 
our reasoning only applies to such systems as are already arbitrarily
complex at the classical level, and are therefore inapplicable 
in the cases of interest, where the map is smooth.
We argue that this is incorrect, for the folowing reason: Since
we are mainly interested in short-range spectral statistics and
since these are dominated by the high iterates of $C_0$, 
we are therefore in fact more interested in $C_0^N$ for $N\gg1$
than in $C_0$ itself. But $C_0^N$ is indeed highly irregular,
since $C_0$ is chaotic. For this reason, it is presumably
qualitatively not different from $C^N$, where $C$ is 
an ``arbitrary'' canonical transformation. While such a 
claim is obviously hard to substantiate, particularly in the 
absence of a measure on the set of canonical transformations,
which would give a precise meaning to the word ``arbitrary''
in the above phrase, it is very reasonable and intuitively appealing.

A related issue arises with respect to integarble and
near-integrable systems: Since sufficiently weak smooth
perturbations of an integrable system has invariant tori which 
cover a non-vanishing measure of phase space, it might appear
that the set of all such systems should have positive measure
in any ordinary sense of the word, in contradiction to 
our claim, which is that almost all systems are fully chaotic. 
This issue is resolved by noting that a generic perturbation
in our sense of the word will, in general, not staisfy the 
smoothness requirements of the KAM theorem, so that any 
amount of such a generic perturbation will lead to completely
irregular dynamics for sufficiently high iterates of the map.

Finally, it remains to see what are the possible special properties
of $C_0$. Since we agree that the only relevant properties
are those that remain invariant under arbitrary iterations of the
map, we are left with very few possibilities. Here are the three
that we are aware of:

\noindent---Time-reversal Invariance: If there exists an anti-canonical
map $T$ (that is, a map which maps the canonical symplectic two-form
on its negative) such that $T^2=1$ and such that
\be
TC_0TC_0=1,
\ee
then the map $C_0$ is said to be time-reversal invariant (TRI). In this
case, we take $\hhh$ to consist of the pairs $(C,TC^{-1}T)$. These
form a group, and it is straightforward to calculate that it leaves the
TRI property invariant. As such, it is the structural invariance group
of an arbitrary $C_0$ which has no specific property apart from TRI.
From this follows that $\Omega$ is simply the set of all TRI bijective
canonical maps, as was in fact to be expected. On the other hand, we
are not (yet!) able to define a measure on $\Omega$, since
neither $\ggg$ nor $\hhh$ have known Haar measures.

This can be readily translated into quantum-mechanical 
language if we note that in the absence of spin, 
$T$ can be represented as complex conjugation in quantum mechanics,
so that $\hhh$ is given by the subgroup consisting of the pairs
of the type $(U, U^t)$, where $U^t$ is the transpose of $U$. 

\noindent---Discrete Symmetries: Let $C_0$ commute with a 
group $G$ of canonical transformations 
$Q$. This means that the evolution given by $C_0$ has symmetries,
given by the maps $Q$. This property is indeed preserved for arbitrary 
iterations of $C_0$. The structural invariance group $\hhh$ 
is then given by the subgroup of all pairs $(C, C^\prime)$
such that both $C$ and $C^\prime$ commute with 
all elements of $G$. Under these 
circumstances, the action of $\hhh$ on $C_0$ respects the property
of commuting with $G$ and generates as a set $\Omega$ the set 
of all bijective canonical maps that commute with $G$. 

Translating into quantum mechanics, this means that the 
quantum version of $\hhh$ consists of all pairs of unitary
matrices that commute with the quantization $U_Q$ of the maps $Q$. 
This means that the unitary matrices acting to the left and
right of $U_C$ can be put in block diagonal form in such a way
that each block belongs to a given irreducible representation
of $G$. Thus, the ensemble generated has a well defined block
structure, in which all blocks statistically independent. On the
other hand, if a given block transforms according to a $d$-dimensional
representation of $G$, the eigenvalues of the corresponding block
will be $d$-fold degenerate. 

\noindent---Systems with mixed dynamics: A related problem arises
with systems having invariant tori, though not, as we shall see,
cantori. Indeed, invariant tori are structures which are preserved
under arbitrary iterations. Furthermore, at least in the most
frequently studied case of two degrees of freedom, they also 
lead to an absolute division of the phase space in disjoint parts. 
The structural invariance group should therefore respect these 
characteristics. The canonical maps belonging to $\hhh$ should
therefore have the same invariant tori as the map $C_0$ iself.
It will therefore automatically also leave the same chaotic
regions invariant. 

Translating this into quantum mechanics, we see that to each torus 
there corresponds a unique wave function via the WKB  prescription.
The above condition therefore means that the group of unitary
maps must act on all these integrable eigenfunctions 
through a phase factor only. From this follows that the eigenphases
in this case can be separated in an integrable spectrum, which consists
of statistically independent eigenphases---since they are acted
upon by the group $U(1)$ with its Haar measure---and a chaotic
part, which can be treated in the usual way described above, and therefore
corresponds to a COE or a CUE for each chaotic region. Note that this 
follows from the preceding remark: Indeed, when two chaotiic regions
are separated by an invariant torus (or an integrable region), 
the index of the chaotic region is a discrete conserved quantity
and acts exactly in the same way as a discrete symmetry. 

This leads to various questions concering the relevance of other
phase space structures, such as homoclinic and heteroclinic
tangles, barriers to transport and cantori. That barriers to 
transport only define separate phase space regions up to a given 
number of iterations is clear. For this reason, we should certainly
not consider it now, since we are limiting ourselves to the simplest,
fully semiclassical case. However, it is clear that such issues will
eventually need to be addressed, since structures of this type
can notoriously cause strong localization effects. While it is
aggreed that these effects are transient when phase space is compact, 
that is the effects disappear at sufficiently small $\hbar$,
they are nevertheless extremely important in practice. Similar 
arguments allow to rule out the other structures as well: Since
they are all invariant sets, it might be argued that they survive 
under arbitrary iteration. This is true as far as it goes, but 
they are complex sets, with fractal structure, which quantum 
mechanics can never fully capture. If we take a finite approximation
to such an invariant set, however, it will become ever more complex
under iteration and therefore, from the point of view of quantum
mechanics, it will eventually become irrelevant. 

\noindent---Non-primitive Period: It is possible that $C_0$ does not
correspond to a primitive period, that is, it might happen
that $C_0$ can be written in the form:
\be
C_0=D^k
\ee
for some $k>1$. In this case, it is not clear to us how to define
a structural invariance group that leaves this property invariant, but the
remedy is clear enough: It is simpler then to quantize
$D$ in the first place. The spectral properties of $U_D^k$ can then
be related to those of $U_{C_0}$. 

We can now go back to the first two properties, and
show that a TRI system will
have a COE, whereas a system with a discrete symmetry will have 
eigenphases thatare divided into statistically independent blocks 
according to the eigenspaces of $U_Q$. To show the first, we point
out that the set $\Omega$ consists of all unitary matrices $U$ such that
\be
U_TUU_T=U^{-1}=U^*.\label{q-TRI}
\ee
Here $U_T$ denotes the anti-unitary map which represents time reversal
in the quantization under consideration. As follows from
well-known theorems, it is always possible to choose a basis 
such that $U_T$ corresponds to complex conjugation. From (\ref{q-TRI})
follows that  $U$ is a symmetric unitary matrix. The action of the
quantum equivalent of $\hhh$ on $\Omega$ is the following
\be
p_U:V\longrightarrow UVU^t,\label{TRI-action}
\ee
where $U^t$ denotes the transpose of $U$ without complex conjugation. 
Clearly, this action leaves the symmetry property of $V$ invariant. 
Equally, if we act on $U_{C_0}$ by means of $p_U$ for all $U$, we
obtain as a set, the set of all symmetric unitary matrices. 
We are therefore led to ask what measure exists on the set of
all symmetric unitary matrices such that it is invariant under
the action of $p_u$ given by (\ref{TRI-action}).
A standard theorem \cite{Cartan} states that
the only such measure is the one known as the Circular Orthogonal
Ensemble (COE). 

Let us now consider the case of discrete symmetry in some
detail, following 
closely the reasoning presented in \cite{Guadalajara,Charles}.
In the case of discrete symmetry, the structural
invariance group $\hhh$ is given by pairs of unitary matrices acting
to the right and to the left of another matrix, which is also 
block diagonal in the same basis. The Haar measure of $\hhh$ is clearly
the product measure of the Haar measures of the various unitary
groups restriced to each block. From this follows that the ensemble
is the product of the various CUE's involved, which is a 
well-known empirical fact, as mentioned in the Introduction. 

Similarly, we had pointed out a difficulty in the case where the
existence of a discrete symmetry was combined with TRI. This can
now be treated in the above manner: We must determine the
structural invariance group $\hhh$ that respects {\it simultaneously\/}
TRI and the symmetry $Q$. In this case, if we describe everything in a 
basis adapted to $G$, it is not true any more that we can describe 
$T$ through simple complex conjugation. Define $U_T$ to be the
{\it anti-unitary\/} representation of $T$ in quantum mechanics.
A moment's thought
shows that two cases are possible: First, $U_T$ leaves invariant
the eigenspaces belonging to the irreducible representations
of $G$ . Then, in this eigenspace, $U_T$
can be represented as complex conjugation and the ensemble
restricted to this eigenspace is indeed the COE, using the same kind of
arguments that lead to the COE when no symmetry is present. 
On the other hand, $U_T$ can act as the combination of complex 
conjugation and the operator interchanging the 
given eigenspace with its time-reversed counterpart.
This is clearly only possible if some $Q$ in $G$ is not itself
TRI, that is, if
\be
TQTQ\neq 1.
\ee
The group $\hhh$ is then given by the pairs $(U,U_TU^{-1}U_T)$. 
In the block diagonal form, the matrix $U$ has entries
\be
\left(
\matrix{&U_1&0\cr
&0&U_2\cr}
\right)
\ee
whereas the matrix $U_TU^{-1}U_T$ gives
\be
\left(
\matrix{
&U_2^t&0\cr
&0&U_1^t\cr}
\right)      
\ee
From this follows readily that the eigenvalues in both eigenspaces
are degenerate, but that the ensemble for each of them is actually
the CUE. If we go through the above considerations from the point
of view of group theory, we see that this phenomenon can
occur whenever some irreducible representation of the
symmetry group is not self-adjoint, that is ,when no basis
can be found in which all the matrices of the representation
are symmetric. The conjugate pair of representations
and their associated invariant subspaces will then be 
interchanged by time-reversal. 

This was verified numerically in \cite{Charles}. 
The specific example considered was a fully chaotic billiard with
threefold symmetry but without any symmetry axis. The rotation 
by $2\pi/3$ is clearly not TRI and it
divides the Hilbert space in three invariant subspaces, generated
by $e^{im\phi}$ for $m$ equal to one, zero or minus one modulo 
three respectively. Whereas the second is clearly invariant
under TRI, the other two are interchanged among each other.
For these, indeed, GUE statistics was clearly observed, whereas in 
the TRI subspace the usual GOE was observed. These observations were later
confirmed and explained on the basis of periodic orbit theory
by Keating and Robbins \cite{Keating&Robbins}.

\section{From eigenphases to eigenvalues}

So far we have only treated the case of a 
bijective canonical map, which can e.g. describe 
the time evolution of a periodically driven system
or a Jung scattering map \cite{Report}. 
An important question, however, is to
transfer this analysis to flows generated by a time-independent Hamiltonian.
The immediate problem is that we must find a way to account for
the fact that the overall density of states is \textit{not\/} given by that
of the corresponding matrix ensembles and can in fact be rather arbitrary,
whereas the fluctuation properties are indeed given by the RMT predictions.
This difficulty was absent in the earlier cases since indeed the density of
states is correctly predicted to be uniform.

To do this, we consider the energy-dependent Poincar\'{e} map $C_E(p,q)$,
where the variables refer to phase space variables in the Poincar\'{e}
surface of section. The Poincar\'e map satisfies the conditions
described above: It is bijective (up to a set of points that start
from the Poincar\'e surface but never return to it. By the
Poincar\'e recurrence theorem, however, these are of measure zero
and can therefore be disregarded). Further, the
part of phase space on which it is defined is a subset of the 
energy surface, and as such is automatically compact for a bounded 
system. We therefore deduce that the eigenphases of
the Poincar\'e map are distributed in the way corresponding
to the symmetries of the map, which themselves correspond to the
symmetries of the system. It now remains to show
in which way this distribution of the eigenphases carries over 
to that of the eigenvalues. 

To this end one proceeds as follows: Bogomolny \cite{bog} has shown that a
semiclassical quantization condition is the following: if $E_n$ is an
eigenvalue, then 
\begin{equation}
\det (\mathbf{1}-T(E_n))=0.
\end{equation}
Thus, if the eigenphases of $T(E)$ are denoted by $\exp (i\phi _j(E))$, then
every time a given $\phi _j(E)$ goes through zero, $E$ is an eigenvalue. It
turns out that the whole procedure is only semiclassical, as the map $T(E)$
is only unitary in the semiclassical limit, but this is not a problem, since
this limit is in any case the only one we are able to handle. Also, outside
of the true semiclassical limit, the relation between chaos and RMT is much
more subtle: in particular, one has problems such as (transient) Anderson
localization, in which chaotic behaviour and randomly distributed
eigenvalues coincide. One also then has to deal with finite tunneling
probabilities and other phenomena associated with the structure of our
canonical map in the complex plane, which we have left out of consideration
entirely, as our understanding of these is still very incomplete.

It therefore appears that we have reduced the problem of determining the
spectrum of a Hamiltonian $H(\widetilde{p},\widetilde{q})$ to the study of
the energy dependence of the eigenphases of the quantized version of its
Poincar\'{e} map. Here the tilde indicate 
canonical variables on the complete phase
space, while canonical variables without tilde
are defined on the surface of section. To handle the problem,
we must first know how the Poincar\'{e} map changes under infinitesimal
changes of $E$. To this end, let us consider two nearby energies
$E$ and $E+\Delta E$. 
The map $C_E$ then maps $(p,q)$
onto $(p^{\prime },q^{\prime })$ and $C_{E+\Delta E}^{-1}$ maps $%
(p^{\prime },q^{\prime })$ onto a point $(P,Q)$ near to the initial
condition. Thus for almost all $(p,q)$

\begin{eqnarray}
P &=&p+\delta p \\
Q &=&q+\delta q,  \nonumber
\end{eqnarray}
where $\delta p$ and $\delta q$ are small. The exceptions occur
if an orbit bifurcates between the two energies $E$ and $E+\Delta E$.
Since this 
transformation is canonical, there must exist a ``Hamiltonian'' 
$\mathcal{T}_E(p,q)$, such that
\begin{eqnarray}
\delta p &=&-\Delta E\frac{\partial \mathcal{T}_E(p,q)}{\partial q} \\
\delta q &=&\Delta E\frac{\partial \mathcal{T}_E(p,q)}{\partial p}  \nonumber
\end{eqnarray}
that is
\begin{eqnarray}
P &=&p-\Delta E\frac{\partial \mathcal{T}_E(p,q)}{\partial q}
\label{canonical} \\
Q &=&q+\Delta E\frac{\partial \mathcal{T}_E(p,q)}{\partial p}.  \nonumber
\end{eqnarray}
Thus the coordinates on the surface of section satisfy the Hamilton
equations for the canonical map $C_{E+\Delta E}^{-1}C_E$ where $\mathcal{T}%
_E(p,q)$ plays the role of the Hamiltonian and $\Delta E$ plays the role of
the time. To show that $\mathcal{T}_E(p,q)$ is the time necessary to return
to the Poincar\'{e} surface if one starts from $(p,q),$ we need to use
reduced action $S_E=\int \{pdq-Hdt\}$. By standard arguments\cite{Landau},
is easy to prove that $-S_E$ is the generating function of $C_E$. Using the
same arguments, $S_{E+\Delta E}$ is the generating function of 
\be
C_{E+\Delta E}^{-1}.
\ee
Finally, the differential of the action of the trajectory that
start on $q$ and ends on $Q$ is given by
\be
dS=d(S_{E+\Delta E}-S_E) 
\ee
or
\be
dS=\frac{dS}{dE}dE. 
\ee
Finally the time to return to the Poincar\'{e} surface of section is given by
\be
\mathcal{T}_E(p_s,q_s)=\frac{dS}{dE}. 
\ee           
Summarising the results for the Poincar\'{e} map, the infinitesimal
canonical transformation close to the identity
\be
C_{E+\Delta E}^{-1}C_E=1-\Delta EC^{-1}(E+\Delta E))\frac d{dE}C(E), 
\ee
means the following: consider the ``Hamiltonian'' $\mathcal{T}_E(p,q)$,
which is defined as the time necessary to return to the surface of section
if one starts from $(p,q)$. This ``Hamiltonian'' generates a flow on the
Poincar\'{e} surface and $C_{E+\Delta E}^{-1}C_E$ is the infinitesimal
canonical transformation corresponding to following this flow for a ``time'' 
$\Delta E$.

If we now follow this through the quantization procedure, we obtain the
following: 
Denote the eigenphases by $\phi _j(E)$ and
$\psi _j(E)$ be the corresponding eigenfunctions. Thus

\be
{\frac{d\phi _j}{dE}=}\langle \psi _j(E)|-iT(E)\frac{dT}{dE}|\psi
_j(E)\rangle 
\ee
             
If we now denote by $\mathcal{H}_E$ the self-adjoint operator corresponding
to $\mathcal{T}_E$, we finally obtain 
\begin{equation}
{\frac{d\phi _j}{dE}}=\langle \psi _j(E)|\mathcal{H}_E|\psi _j(E)\rangle .
\end{equation}
Now we must make some key approximations: First, we remember that we are in
the semiclassical limit, that is, that the classical function $\mathcal{T}%
_E(p_s,q_s)$ is smooth compared with the $\psi _j(E)$. This implies that
instead of using, say, the Wigner distribution in computing the l.h.s. of
Eq.~(8), we can use a smoothed version such as the Husimi distribution
without great error. Since the $\psi _j(E)$ are eigenfunctions of a matrix
representing a totally structureless map, their Husimi distributions will be
spread uniformly all over phase space. This would follow from our
considerations on structural invariance, but is equally confirmed by a
rigorous theorem of Shnirelman's concerning eigenfunctions of chaotic
systems. From this one finally gets 
\begin{equation}
{\frac{d\phi _j}{dE}}=\overline{\mathcal{T}_E(p,q)},  \label{same-velocity}
\end{equation}
where the overline denotes average over phase space. The crucial points to
note is that the right-hand side depends on a classical energy-dependent
quantity. However, it is quite independent of
the specific wave function and hence of $j$. 
Since there are $N$ eigenphases on the unit circle and they
move with a velocity on the
order of one, they will cross zero at energies which differ by an order of 
$1/N$. This is natural, since we have chosen our scale of 
energies to be the
classical one. Therefore, from one eigenvalue to the next, 
the velocity at
which the eigenphases moves hardly changes. Thus, locally at least, 
the motion of the eigenphases is quite rigid. This means that the 
RMT properties of the eigenphases translate 
directly into corresponding properties of the eigenvalues of the 
system. On the other hand, it is important to realize that
this constancy does not hold forever. Two effects
will eventually alter change the velocity: first, the average on the
l.h.s. of Eq.~(\ref{same-velocity})
will experience a secular change as $E$
changes. This corresponds to the secular change in the density of states which
is usually eliminated by unfolding the spectrum. On the other hand, another
effect may well come into play even before this secular change becomes
noticeable. The point is that Eq.~(\ref{same-velocity})
is only true as a statistical statement
and there are fluctuations around the mean velocity. The most obvious 
cause for such fluctuations are departures of the Husimi distribution from
equidistribution. Such deviations are well-known to exist, namely the
so-called ``scars'' near short periodic orbits. It could therefore well be
that these accumulated fluctuations account for some of the effects
due to short periodic orbits. To explore this possibility, however, we would
presumably require an understanding of scars which we do not have at
present. Note also that the long-range stiffness found by Berry
\cite{berry} is a phenomenon that lies beyond the range of validity
of the above remarks. Indeed, its onset is at an infinite distance
in terms of the mean level spacing, as we go to the semiclassical
limit. 

\section{The quantum Poincar\'{e} map on the billiards}

Here we will give a numerical example of the principal results of the last
section by studying the properties of the eigenphases for the rectangle and
Bunimovich sttadium with Neumann boundary conditions. 
The latter is defined as the (convex) region enclosed by two 
semicircles of radius $R$ connected by two parallel segments of
length $L$. This system has been shown \cite{buni17} to be completely
chaotic.     
Following Bogomolny 
\cite{bog} and Boasman \cite{boasman}, we will use the boundary integral
method for the Helmholtz equation. For a billiard 
defined by a region $A$ with boundary $B$ it is
\begin{eqnarray}
-\frac{\hbar ^2}{2m}\nabla ^2\Psi _n(x)&=&E_n\Psi _n(x) \quad x\hbox{ inside } 
A \\ 
\frac{\partial \Psi _n(x)}{\partial n}&=&0 \qquad x\hbox{ on }B, \nonumber
\end{eqnarray}
where $\frac \partial {\partial n}$ is the normal derivative to the boundary
at point $x$. Introducing the usual Green function $G_0(x^{\prime },x;E)$
for the free Laplacian and using the Green identity, on has after
integration over the complete region,
\begin{eqnarray}
\frac 12\Psi_n(q^{\prime })&=&
\frac{\hbar^2}{2m}\oint_Bdq\frac{\partial G_0(q^{\prime },q;E)}{\partial n}%
\Psi_n(q)- \\ \nonumber
&-& \frac{\partial \Psi _n(q)}{\partial n}G_0(q^{\prime
},q;E)+\\  \nonumber
&+&(E-E_n)\int_AG_0(q^{\prime },x;E)\Psi _n(x)dx,
\end{eqnarray}
where $q$ and $q^{\prime }$ denote coordinates on the boundary $B$.
             
Setting $E=E_n$ and imposing the Neumann boundary conditions (other kind of
conditions are straightforward) the last equation takes the form:

\begin{equation}
\frac{\hbar ^2}m\oint_B\frac{\partial G_0(q^{\prime },q;E)}{\partial n}\Psi
_n(q)dq=\Psi _n(q^{\prime }).
\end{equation}
Discretising the boundary by $N$ equally spaced points $\{q\}_{i,j=1,\dots
,N}$ and approximating the integral by the trapeze rule, the last equation
takes the form 
\begin{equation}
\frac{\hbar ^2\Delta }m\sum_{j=1}^N\frac{\partial G_0(q_i,q_j;E)}{\partial n}%
\Psi _n(q_j)=\Psi _n(q_i)
\end{equation}
and the eigenvalues of the billiard can be obtained by looking for the roots 
\begin{equation}
\det (1-dT_{i,j}(E))=0\qquad i,j=1,\dots ,N
\end{equation}
when $E$ is varied as parameter. Here $d$ is the distance introduced by the
discretisation of the integral and 
\be
T_{i,j}(E)=\frac k{2i}\left\{ \widehat{n}(q)\cdot (q_s-q_t)\right\}
H_1(k|q_s-q_t). 
\ee
The Hankel function is used because in two dimensions 
\begin{equation}
G_0(q_i,q_j;E)=\frac{-i}4H_0^{(1)}\left( \sqrt{\frac{2mE}{\hbar ^2}}%
|q_i-q_j|\right) ,
\end{equation}
and $\hat{n}(q_i)$ is the outgoing normal to the boundary at point $q_i$.
To look for the eigenphases we now solve the eigenvalue
equation for the matrix $T_{i,j}(E)$.

We begin by studying the properties of the eigenphases of the quantum
Poincar\'{e} map in a rectangle and in a quarter of stadium. The quantum
surface of section that we used is the boundary of the billiard. The
eigenphases of $T_{i,j}(E)$ for 
a given value of the energy are shown in
Fig.~1. The unitarity of the QPM is better in the high energy regime as
expected. Also the norms are closer to one when the respective phase is zero
or $\pi $. This behaviour is shown in Fig.~2 where we plotted the norms
as function of the phase.

To study in a detailed way the behaviour of the eigenphases of Fig. 1 as
functions of the energy we plot in Fig.~3 their phases and norms as function
of the energy. This figure suggest that the phases for the stadium move in a
``rigid way'' when the energy is varied whereas for the rectangle the phases
move at least with two velocities. To quantify the velocities we plotted in
Fig. 4a the distribution of velocities
for the stadium. From this figure is easy to conclude
that all eigenphases move basically with the same velocity for the stadium.
These results shown the validity of Eq. \ref{same-velocity} 
(this conjecture was also made
by Prange \cite{prange}) that the eigenvalues in the chaotic case
move in a rigid way independently of the number of eigenvalue. This is not
the case for the rectangle (integrable system) as we can see in Fig.~4b.

To study the secular variation of eigenphases we plot in Fig. 5 the
distributions of the velocities for the stadium for 2 
different ranges of $k$. Also, in table I we tabulate the mean values as
well as the widths of these and other velocity distributions. From this it is
possible to see that these widths are basically constant when $k$ is varied 
in the chaotic case.
This is accord with the theoretical prediction of Doron and Smilansky for a 
$k^{-1}$ behaviour of the width 
as function of $E$. We also show the widths for integrable systems in table
II. It is seen that, for comparable values of the mean, the widths are 
significantly larger than in the chaotic case. Again, the widths are
found to be independent of $k$. The fact that the widts are larger in the
integrable case is therefore quite robust under variations of the energy.

The numerical results of this section thus confirm the theoretical
predictions: for the completely chaotic case the eigenphases move almost
with the same velocity and thus its statistical properties translate to
those of the energies. Nevertheless this conclusion does not hold forever.
When the eigenvalues have given a complete turn on the unit circle, large
correlations must occur. These long range correlations are well known: the
saturation in the spectral rigidity (equivalently in the Dyson-Mehta
statistics) related to the shortest periodic orbit\cite{berry}, and the
periodicity in the Fourier transform of the two levels cluster function, 
$b_2(t)$ for Rydberg molecules \cite{lombardi}. Also eigenphases with small
velocities have been found in the stadium. These correspond to the bouncing
ball states. Note finally that the techniques proposed by Lombardi
and coworkers \cite{lombardi} for Rydberg molecules in the framework
of Multichannel Quantum Defect Theory can be interpreted from the
same point of view and provide an exactly unitary Quantum Poincar\'e
map. These maps dispaly similar properties as those described above 
for the Poincar\'e map of the stadium. In particular, the stiffness
in the motion of the eigenphases for the case of chaotic
motion is also observed in this system \cite{rafael}.

\section{Conclusions}

We have presented a systematic approach to the connection between RMT and
individual dynamical systems. This connection is in a sense of a probabilistic
type: it rests basically on conclusions of the form: 
let a given system be a
typical representative of a certain ensemble: If
the elements of this ensemble have
a given property with probability one, the original system 
presumably also has the
stated property. While this line of argument is wide open to 
criticisms from the mathematical side, it is undeniably useful 
at the heuristic
level. Further, similar lines of reasoning are frequently used in
physics---for example in statistical mechanics---when trying to
apply an ensemble description to an individual
system. A more genuine concern concerns the construction of the ensembles:
as we are not able to construct ensembles on the set of all canonical
transformations, we must first construct a set of canonical transformations
at the purely classical level and then translate this into quantum mechanics
in order to obtain a reasonable candidate for a measure. On the other 
hand, it may be possible---and we would like to suggest that 
this would be important---to find an invariant measure
on the set of all bijective canonical transformations
on a compact phase space. If this were possible, we could 
define natural classical equivalents to the CUE and the COE.
One could then argue in an entirely classical way that a given map
with chaotic dynamics is a typical element of such a classical ensemble
of canonical maps. Its quantization would therefore belong to the associated
quantum ensemble, whence the required spectral properties 
follow. 

Another limitation is the purely semiclassical nature of the analysis.
It does not, for example, apply to
quantum localization: as long as the phase space $\Gamma $ is compact,
localization is only transient, and therefore outside the immediate range of
application of semiclassics. Non-compact phase spaces, on the other hand,
also present problems relating to the existence of an invariant measure,
not only in the classical, but also in the quantum case.

On the other hand the translation of the fluctuation properties from the
eigenphases to the energy spectrum is particularly transparent. Thus the
fluctuation properties of the energies are the same as those of the Gaussian
ensembles with probability one. This approach shows that not only do the RMT
predictions hold for the two-point function at intermediate energy
distances, but that they should hold at all energy scales and for all
correlation functions. Also the saturation of the long rage stiffness for
the energy spectrum is understood in terms of complete turns of the
eigenphases on the unit circle.  Finally we want to mention that
the properties of the velocities of the eigenphases promise to be a 
new quantum signature of chaos.

\section{Acknowledgments}

This work is supported by CONACYT-M\'exico and by DGAPA-UNAM.

\begin{figure}[tbp]
{\hspace*{-.5cm}\psfig{figure=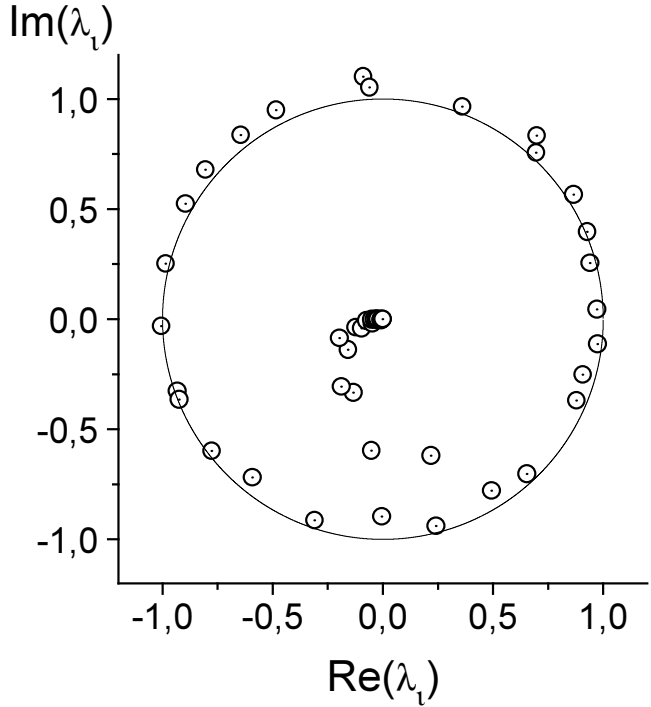,height=10cm,angle=-90}} \vspace*{.12in}
\caption{Eigenvalues of the operator associated to the quantum Poincar\'{e} 
map for a quarter of stadium of size $L=20$ y $R=40$. We have taken $N=58$ 
and $k=0.49$. Note that the unitarity is better for phases close to $0$ and 
$\pi$.} \label {fig:1}
\end{figure}

\begin{figure}[tbp]
{\hspace*{-.5cm}\psfig{figure=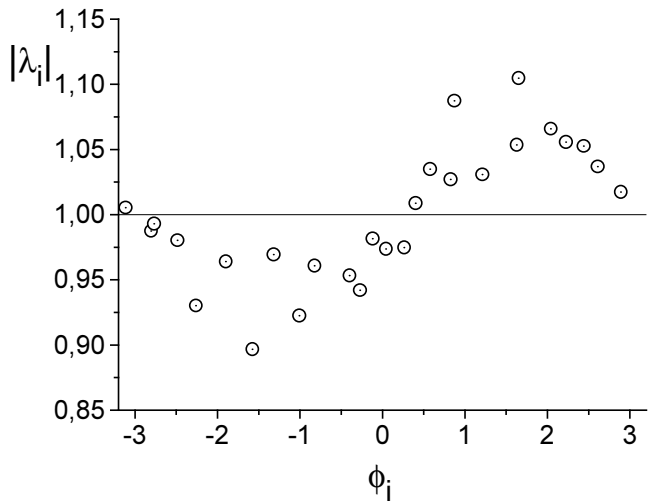,height=10cm,angle=-90}} \vspace*{.12in}
\caption{Norms of the eigenvalues of Fig. 1 as a function of their phases. We 
plotted only those with a norm larger than $0.8$.} \label {fig:2}
\end{figure}

\begin{figure}[tbp]
{\hspace*{-.5cm}\psfig{figure=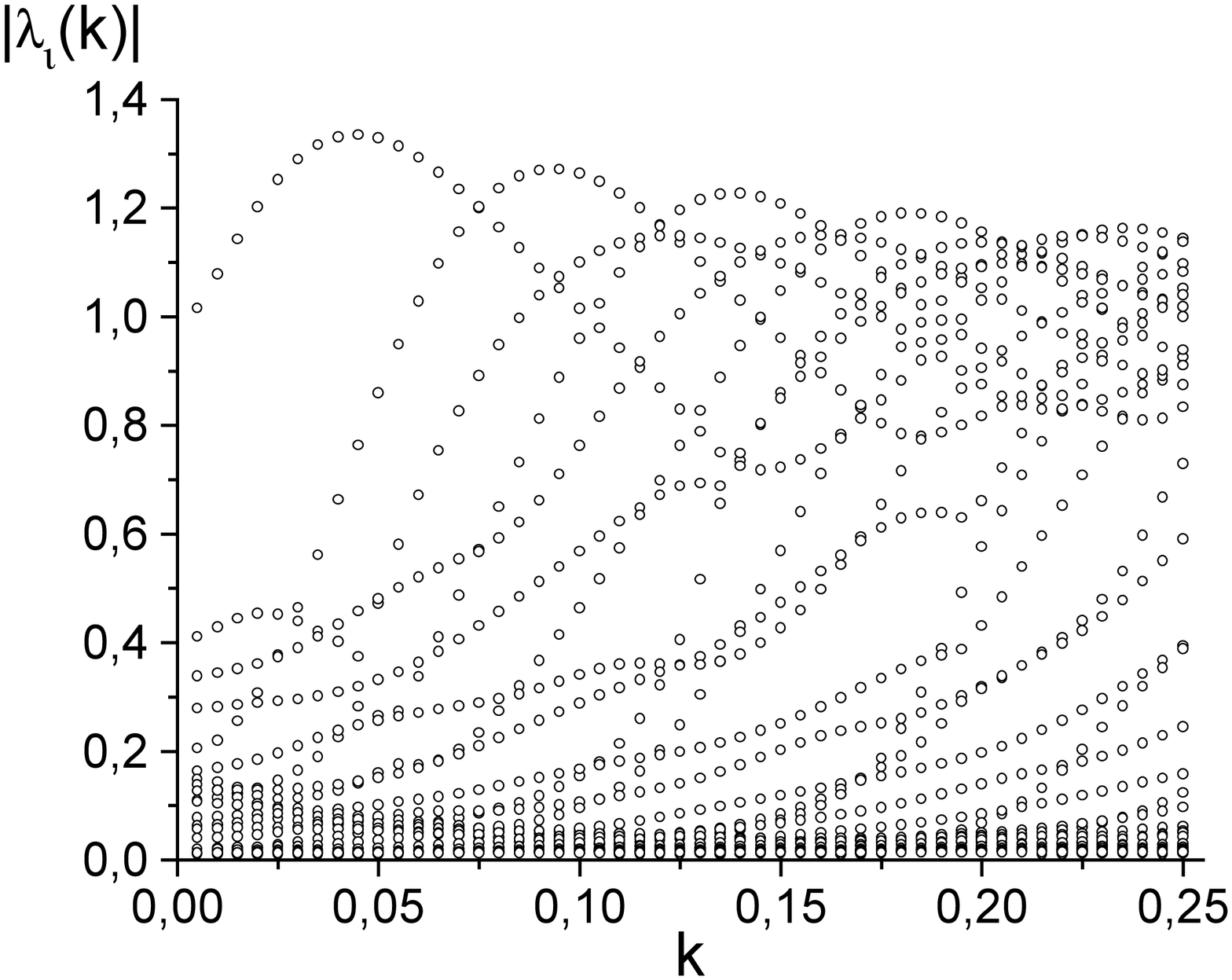,height=9cm,width=5cm,angle=-90}} 
{\hspace*{-.5cm}\psfig{figure=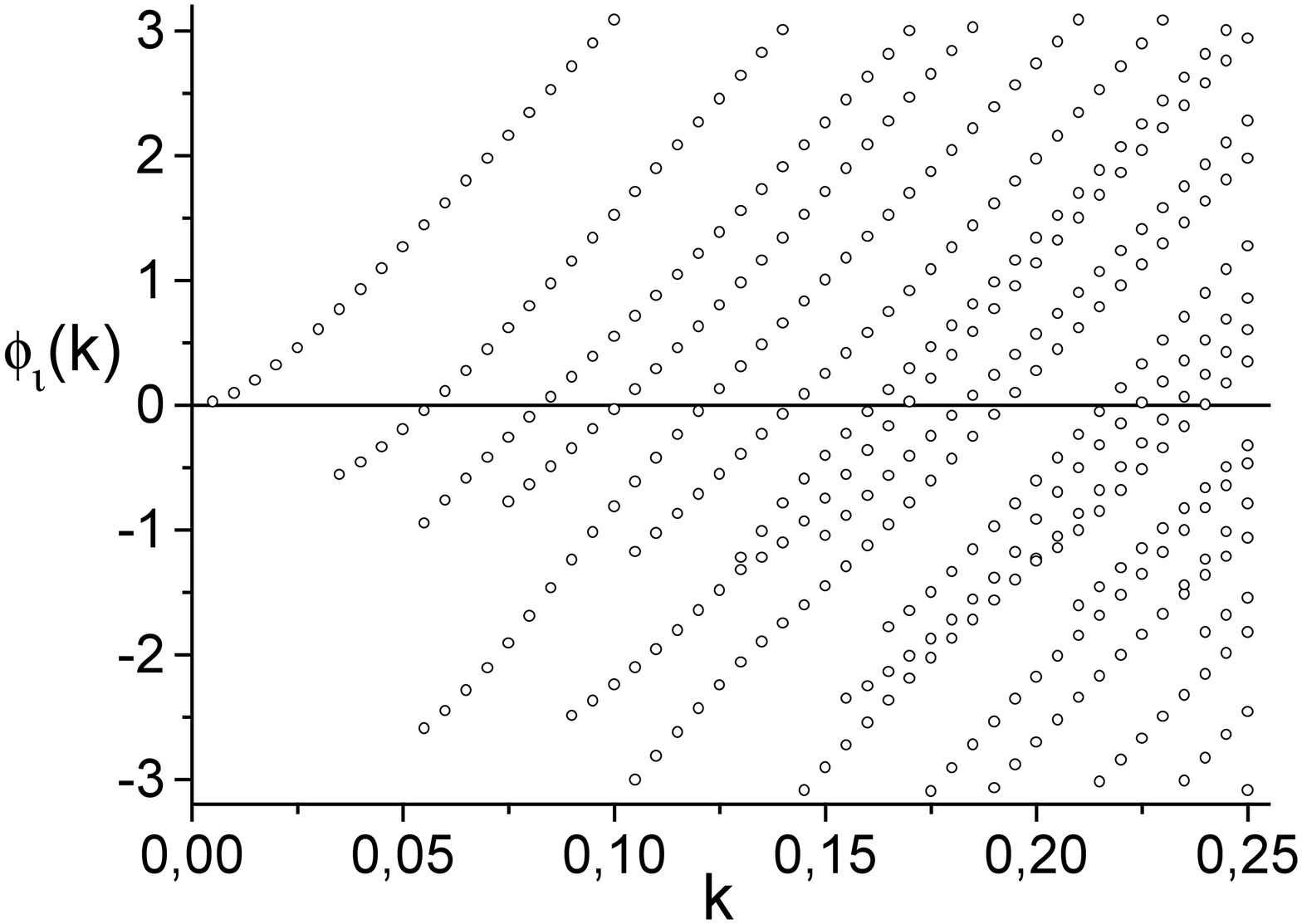,height=9cm,width=5cm,angle=-90}} 
\caption{a) Norms and b) phases of the eigenphases of the 
quantum Poincar\'e map for the same stadium for $N=58$ as 
a function of $k$.} \label {fig:3}
\end{figure}

\begin{figure}[tbp]
\vspace*{-.7cm}
{\hspace*{-.5cm}\psfig{figure=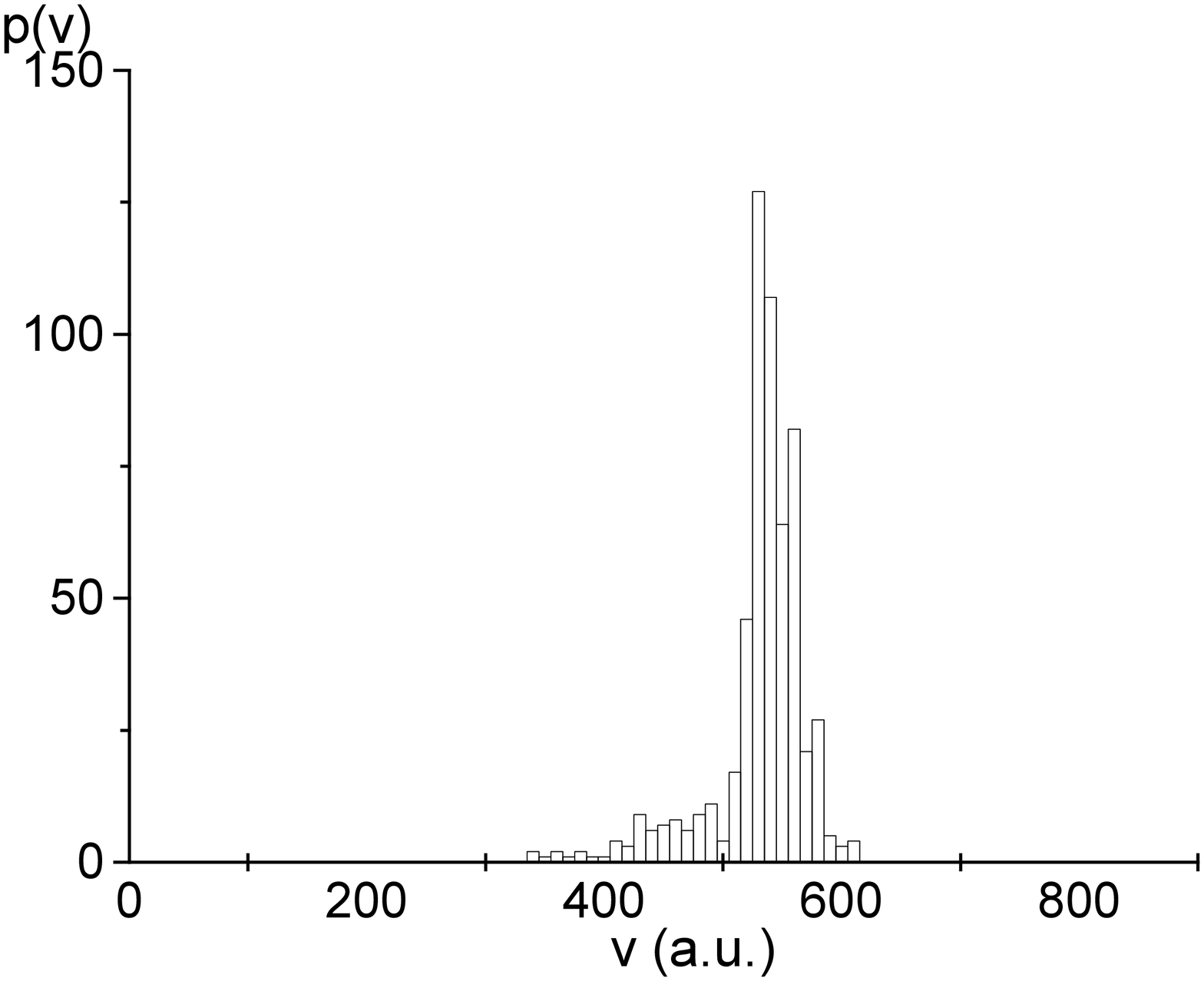,height=9cm,width=5cm,angle=-90}} 
{\hspace*{-.5cm}\psfig{figure=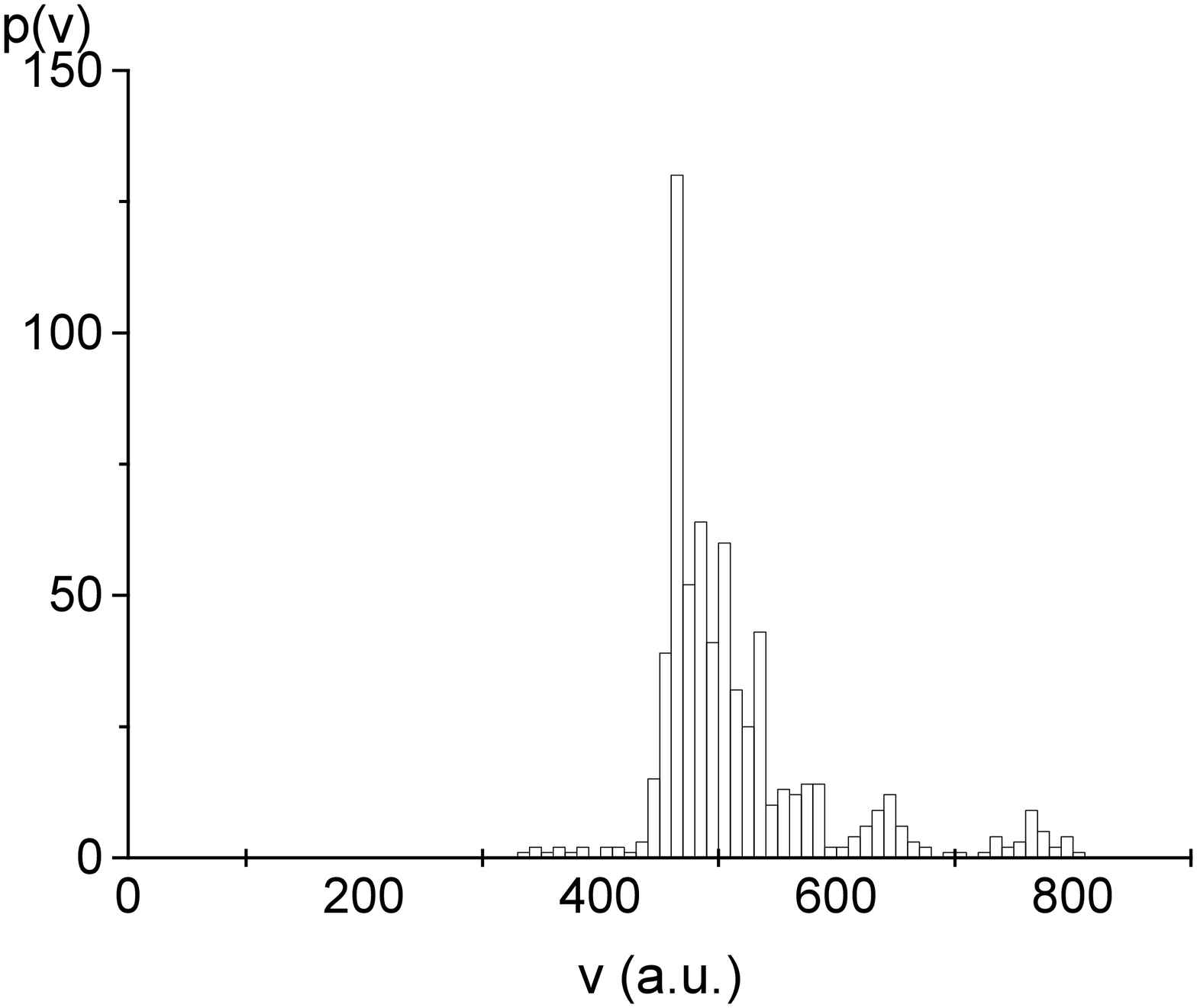,height=9cm,width=5cm,angle=-90}} 
\vspace*{.12in}
\caption{Velocity distribution for a) the same quarter of stadium for 
$k \sim 0.45$, $N=290$ and b) the
rectangle with an aspect ratio given by the golden 
mean. We use $N=300$. We only
used the eigenphases whose norm is greater than $0.6$.} \label {fig:4}
\end{figure}

\begin{figure}[tbp]
{\hspace*{-.5cm}\psfig{figure=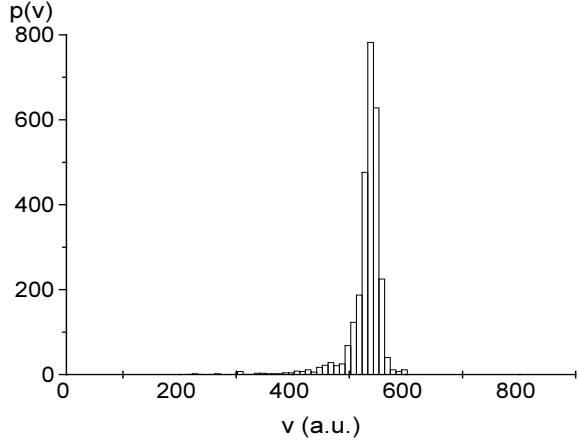,height=9cm,width=6cm,angle=-90}} 
\vspace*{.12in}
{\hspace*{-.5cm}\psfig{figure=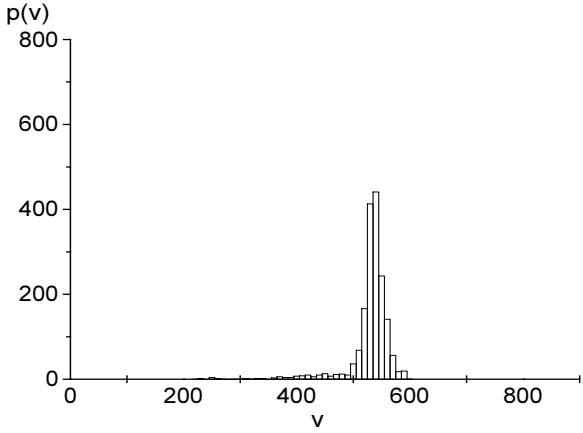,height=9cm,width=6cm,angle=-90}} 
\vspace*{.12in}
\caption{Velocity distributions for the same quarter of stadium with $N=290$ and
a) $k\sim 0.02$ and b) $k\sim 0.09$.} \label {fig:5}
\end{figure}

%
%
%
%

{\begin{table}
\begin{tabular}{|c|c|c|} \hline
$k$ & Center & Width \\ \hline 
0.02 & 530.586 & 39.839 \\
0.03 & 528.960 & 39.140 \\
0.04 & 528.905 & 42.797 \\
0.05 & 528.826 & 42.575 \\
0.06 & 529.078 & 39.405 \\
0.07 & 529.008 & 36.684 \\
0.08 & 529.788 & 37.352 \\
0.09 & 530.005 & 33.713 \\ \hline
\end{tabular}
\caption{Centers and widths of the velocity distributions of the same quarter
of stadium for various values of $k$ and $N=290$.}
\end{table}
}

\begin{table}
\begin{tabular}{|c|c|c|} \hline
$k$ & Center & Width \\ \hline 
0.02 & 515.438 & 77.142 \\
0.03 & 516.801 & 85.977 \\
0.04 & 519.923 & 82.403 \\
0.05 & 519.559 & 78.797 \\
0.06 & 520.720 & 73.579 \\
0.07 & 521.260 & 72.360 \\
0.08 & 522.025 & 72.095 \\
0.09 & 522.057 & 71.590 \\ \hline
\end{tabular}
\caption{Centers and widths of the velocity distributions of the same 
rectangle for various values of $k$ and $N=300$.}
\end{table}

\end{document}